\documentclass[12pt]{article}
\usepackage{amsmath,amssymb}
\usepackage{graphicx}
\usepackage{float}
\usepackage{hyperref}
\begin{document}
\title{Optimal Execution Strategies Incorporating Internal Liquidity Through Market Making}

\author{
Yusuke MORIMOTO
\thanks{
MUFG Bank, Ltd, 1-4-5 Marunouchi, Chiyoda-ku, Tokyo 100-8388, Japan, \
E-mail: yuusuke.morimoto@gmail.com} }
\date{}
\maketitle
\begin{abstract}
This paper introduces a new algorithmic execution model that integrates interbank limit and market orders with internal liquidity generated through market making. 
Based on the Cartea et al.\cite{cartea2015algorithmic} framework, we incorporate market impact in interbank orders while excluding it for internal market-making transactions. 
Our model aims to optimize the balance between interbank and internal liquidity, reducing market impact and improving execution efficiency.
\end{abstract}


\section{Introduction}
When executing large orders, carrying out trades over a short period can trigger unfavorable market movements, known as market impact. 
On the other hand, spreading the execution over a longer time exposes the trader to market risk. 
Algorithmic execution strategies must navigate this trade-off, 
and such strategies are generally categorized into static strategies, 
which follow a predetermined execution schedule, and dynamic strategies, 
which adjust the schedule in response to market conditions.

The seminal work of Almgren and Chriss \cite{almgren2001optimal} and Bertsimas and Lo \cite{bertsimas1998optimal} introduced a foundational model in this field, 
representing the market mid-price as a continuous diffusion process with market impact 
incorporated as a drift term. This model provides a framework for deriving static optimal execution strategies using variational methods.
Regarding static strategy, many studies have also extended the modeling of market impact structures, 
including nonlinear market impact and resilience and transient impact in Dang\cite{dang2017optimal}, Gatheral et al. \cite{gatheral2011exponential}, Gatheral and Schied \cite{gatheral2012transient}, and Curato\cite{curato2017optimal}.

In real markets, it is critical to account for bid-offer spreads and to optimally balance the use of market orders(MOs) and limit orders(LOs). 
Cartea et al.\cite{cartea2015algorithmic}, Cartea and Jaimungal \cite{cartea2015optimal} extended this framework by modeling LOs using a jump-diffusion process incorporating Poisson arrivals, while representing MOs as interventions within an impulse control framework. 
They solved the associated Hamilton-Jacobi-Bellman Quasi-Variational Inequality (HJB QVI) to derive dynamic strategies.

In financial institutions, market making(MM) plays a important role in providing liquidity to the market by simultaneously quoting buy and sell orders and acting as a counterparty to other participants. 
Algorithmic execution can leverage internal liquidity through market making, attracting order flow by offering more favorable prices. 
This approach, particularly in OTC markets such as the FX market, reduces market impact and enables efficient execution compared to trading directly in external markets.
Moreover, this mechanism benefits both the institution’s algorithmic execution clients and the clients receiving quotes on the market-making side. 
As such, it has become a widely adopted practice in the industry.

There has been extensive research on market making in Avellandeda and Stoikov\cite{avellaneda2008high}, Gu\'eant et al.\cite{gueant2013dealing} and Cartea et al.\cite{cartea2017algorithmic}.
All of which focuses on determining optimal strategies for market makers. 
This paper, however, examines optimal strategies from the perspective of an algorithmic execution agent, 
rather than considering the profitability of market makers. 
Since this method essentially transfers the flow to algorithmic execution clients, 
it does not generate profit and loss (PL) for market makers. 

This paper proposes a novel model that incorporates not only interbank MOs and LOs but also internal liquidity generated through MM as potential execution options. 
The motivation for this approach lies in the ability to quantify the optimal extent of favorable pricing in market making and the ideal integration of interbank and internal liquidity during execution.
Our model builds upon the Cartea et al.\cite{cartea2015algorithmic} framework by extending it to account for market impact in interbank LOs and MOs, while excluding market impact for internal orders executed through MMs. 
This distinction allows us to analyze the interaction between interbank and internal liquidity. 
We derive the corresponding HJB QVI, solve it numerically, and analyze the resulting strategies to shed light on their characteristics.

\section{Utilization of liquidity in the market making}
This chapter provides a detailed explanation of algorithmic execution utilizing market making. 
Financial institutions operate both as market makers and takers. 
Specifically, as market makers, they primarily supply liquidity to client markets, and the resulting positions are hedged in the interbank market as market takers. 
Additionally, algorithmic execution services provided by financial institutions typically involve acting as a taker to execute large client orders on their behalf.

Such algorithmic execution is generally carried out through limit orders (LO) or market orders (MO) in the interbank market. 
If the institution is also engaged in market making, it can attract necessary flow by offering competitive prices to its market-making clients. 
This flow can then be handed over to the algorithmic execution side, 
creating mutually beneficial transactions for both the clients using algorithmic execution services and the clients engaging with the institution's market-making services. 
This integrated approach enhances efficiency and value for all parties involved.

\begin{figure}[]
  \centering
  \includegraphics[width=0.9\textwidth]{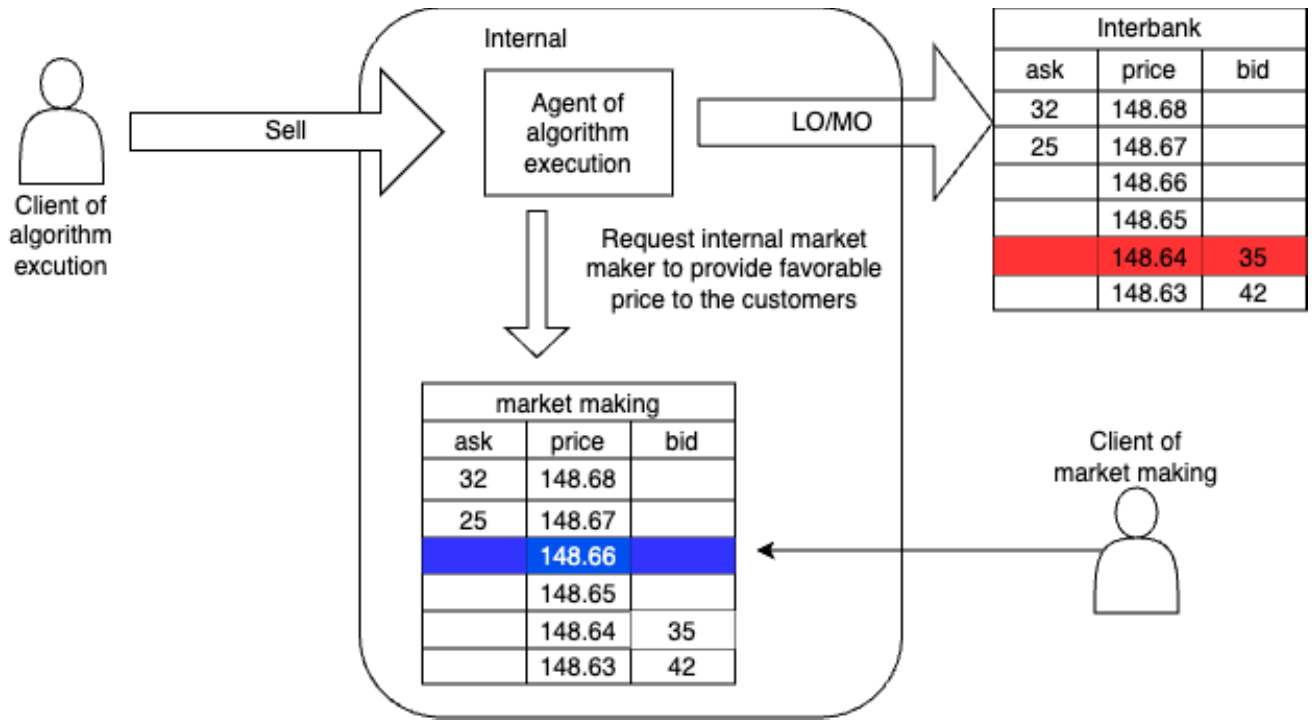}
  \caption{Image of algorithm execution utilizing MM}
  \label{fig:mm}  
\end{figure}

Figure \ref{fig:mm} shows an illustration of algorithmic execution utilizing market making.
\begin{itemize}
  \item FX market image (OTC), USD/JPY.
  \item Client order: Sell algorithmic execution.
  \item Agent of alogorithmic execution requests internal market maker to provide favorable price to the client of market making side.
  \item For clients on the maker side, they can purchase at a special discounted price, which is lower than the usual ask price.
  \item For clients utilizing algorithmic execution, they can sell at a price higher than the interbank bid price while also minimizing market impact.
\end{itemize}
\begin{center}
\end{center}

\section{Model}\label{sec:model}
\subsection{Mathematical Framework: Combined Stochastic Control and Impulse Control}\label{sec:math}
In this section, we follow \O{}ksendal and Sulem\cite{oksendal2019stochastic} and Seydel\cite{seydel2009existence} for more rigorous proofs.
Let $(\Omega, \mathcal{F},\{\mathcal{F}_t\}_{t\in [0, T]}, P)$ be a filtered probability space.
Let $A$ be a compact non-empty metric space.  Control process $v(t, \omega): [0, T] \times \Omega \rightarrow A$ is cadlag $\{\mathcal{F}_{t}\}$-adapted.
$\{W_t\}$ is $\{\mathcal{F}_{t}\}$-adapted $d_1$-dimensional Brownian motion and $\{N_t\}$ is $\{\mathcal{F}_{t}\}$-adapted $d_2$-dimensional poisson process with intensity $\lambda^v$.
Let $\theta = \{(\tau_i, \zeta_i)\}_{i\ge0}$, such that $\tau_0 \le \tau_1 \le \tau_2 \le \cdots$ be $\{\mathcal{F}_t\}$-stopping times and $\zeta_i$ be $\mathcal{F}_{\tau_i}$-measurable $\mathcal{Z}\subset {\bf R}^k$-valued random variable, $i=1, 2,\cdots.$ 
The $k$-dimensional state process $Y$  follows the stochastic differential equation with impulses
\begin{align}\label{eq:sde}
  &dY_t = \mu(Y_{t-}, v_{t-})dt + \sigma(Y_{t-}, v_{t-}) dW_t + \gamma(Y_{t-}, v_{t-})dN_t, t \in (\tau_{i-1}, \tau_{i}), \\
  &Y(\tau_i) = \Gamma(\check{Y}(\tau_i-), \zeta_i), i = 1, 2, \cdots, 
\end{align}
where the term $\check{Y}_{\tau_i-}$ denotes the value of the controlled $Y$ in ${\tau_i}$ with a possible jump of the
stochastic process, but without the impulse, i.e., $Y_{\tau_i-} + \Delta Y_{\tau_i}$, \ $\Gamma: {\bf R}^k \times \mathcal{Z} \rightarrow {\bf R}^k, \mu: {\bf R}^k \times A \rightarrow {\bf R}^k, \sigma: {\bf R}^k \times A \rightarrow {\bf R}^{k\times d_1}$ and $\gamma: {\bf R}^k \times A \rightarrow {\bf R}^{k\times d_2}$
satisfying the necessary conditions such that existence and uniqueness of the SDE (\ref{eq:sde}) is guaranteed.

Let $\mathcal{A}(s, y)$ be a class of combined stochastic control $v: [s, T]\times \Omega \rightarrow A$ and impulse controls $\theta$ such that 
existence and uniqueness of the SDE (1) be guaranteed. 

$$ \mathcal{M}\phi (y) = \sup\{\phi(\Gamma(y, \zeta)); \zeta \in \mathcal{Z}\}.$$
$$L^v \phi(y)=\mu(y,v)\partial_y \phi(y)+\frac{1}{2}\sigma(y,v)^2\partial_{yy}\phi(y)+(\phi(y+\gamma(y,v))-\phi(y))\lambda^v.$$

Then we consider the following value function
\begin{align} \label{eq:valGen}
 H(t, y) = \sup_{(v, \theta) \in \mathcal{A}(t, y)} E\left[g(Y_T) + \int_t^T f(s, Y_s, v_s)ds |\mathcal{F}_t \right] , 
\end{align}
where $f: {\bf R}^+\times {\bf R}^k \times A \rightarrow {\bf R}, g: {\bf R}^k \times A \rightarrow {\bf R}$ are measurable.
The value function (\ref{eq:valGen}) is a unique viscosity solution of the following HJB QVI(Hamilton-Jacobi-Bellman Quasi Variational Inequality)
\begin{align}\label{HJBQVI}
 &\max \left\{\partial_tH(t,y) + \sup_{v\in A}(L^vH(t,y) + f(t, y, v)), 
 \mathcal{M}H(t, y) - H(t, y)\right\}=0,\\
 &H(T, y) = g(y).\nonumber
\end{align}

\subsection{The Model Utilizing the Liquidity of Internal Market Making 
with External Limit and Market Orders}
We extend the Limit and Market Order model in Cartea et al. \cite{cartea2015algorithmic}. 
The setting involves selling a fixed notional amount $Q_0$ of underlying asset(a similar setup applies to buying).
\begin{itemize}
  \item $T$ is the terminal time at which the liquidation ends.
  \item $X = (X_t)_{0\le t \le T}$ denotes the agent's cash process with $X_0 = 0$.
  \item $S = (S_t)_{0\le t \le T}$ is the asset's mid price with $S_0 > 0.$
  \item $Q = (Q_t)_{0\le t \le T}$ is the agent's inventory which remains to be liquidated.
  \item $\delta^L = (\delta^L_t)_{0\le t \le T}$ denotes the depth at which the agent posts limit sell orders, i.e. 
  the agent posts LOs at a price of $S_t + \delta^L_t$ at time $t$.
  \item $\delta^I = (\delta^I_t)_{0\le t \le T}$ denotes the ask spread at which the agent shows through the market making, i.e. 
  the agenet requests MM to provide price of $S_t + \delta^I_t$ at time $t$.
  \item $N^L = (N^L_t)_{0\le t \le T}$ denotes the counting process corresponding to the number of MOs that matched the agent's LOs.
  \item $N^I = (N^I_t)_{0\le t \le T}$ denotes the counting process corresponding to the number of internal orders obtained through MM.
  \item $\lambda^L = (\lambda^L_t)_{0\le t \le T}$ denotes the intensity of $N^L$. $\lambda^L_t = \lambda_L e^{-\kappa_L \delta^L_t}$ with parameter $\lambda_L > 0$ and $\kappa_L > 0$. 
  The larger $\delta^L$ (the higher the selling LO places), the lower intensity $\lambda^L_t$ that means the lower probability of a match for LO.
  \item $\lambda^I = (\lambda^I_t)_{0\le t \le T}$ denotes the intensity of $N^I$. $\lambda^I_t = \lambda_I e^{-\kappa_I \delta^I_t}$ with parameter $\lambda_I > 0$ and $\kappa_I > 0$.
  The larger $\delta^I$ (the higher the price of MM provide), the lower intensity $\lambda^I_t$ that means the lower probability of a match through MM.
  \item $\theta = \{(\tau_i, \zeta_i)\}_{i\ge 1}$ denotes that $\tau_i$ is a stopping time that represents the agent's MO timing and 
  $\zeta_i$ is a ${\bf N}$-valued $\mathcal{F}_{\tau_i}$ measurable random variable that denotes the size of MO, i.e. we assume the minimum unit of the trade amount is $1$.
  We formally assume $\tau_0 = 0.$
  \item $\xi$ denotes the bid spred, i.e. the distance between mid and the best bid. We assume it is deterministic.
  \item $\alpha$ denotes the parameter for the additional market impact of a market order at the terminal date  $T$.

\end{itemize}
The dynamics of the state state process $Y_t = (X_t, S_t, Q_t)$ is as follows.
\begin{align}
  &dX_t = (S_t + \delta^L_{t-} -\alpha_L\lambda^L_{t-}) dN^{L}_t +(S_t+\delta^I_{t-})dN^{I}_t, \label{eq:dX}\\
  &dS_t = \sigma dW_t, \label{eq:dS}\\
  &dQ_t = -dN_t^{L} -dN^{I}_t, \label{eq:dQ} 
\end{align}
for $t \in (\tau_{i-1}, \tau_i),$ and $Y_{\tau_i}$ is
\begin{align}
  &X_{\tau_i} = \check{X}_{\tau_i-} + (S_{\tau_i}-\xi) \zeta_i -\alpha_M \zeta_i^{{\beta_M}}, \label{eq:X}\\
  &S_{\tau_i} = S_{\tau_i-}, \label{eq:S}\\
  &Q_{\tau_i} = \check{Q}_{\tau_i-} -  \zeta_i, \label{eq:Q}
\end{align}
for $i = 1, 2, \cdots.$

The equations (\ref{eq:dX})-(\ref{eq:dQ}) represents the dynamics at the moment when orders from LO and MM match.
\begin{itemize}
  \item The first term of (\ref{eq:dX}) represents the increase in cash when the LO placed at $S_t + \delta^L_t$ matches. 
  Here, $-\alpha^L\lambda^L_t$ represents the market impact that is assumed to be a linear of the intensity of the Poisson process that represents the number of LO matches. 
  This is analogous that many models incorporate market impact in the form of linear function of $\frac{dQ}{dt}$
  that cannot be defined in this model.
  \item The second term of (\ref{eq:dX}) represents the increase in cash when a match occurs at the price of $S_t + \delta^I_t$ through MM.
  The key point of this model is that market impact is not incorporated for orders executed through MM. 
  This differentiates them from interbank orders and reflects what is actually observed in real markets, particularly in OTC markets such as the FX market.
  \item Equation (\ref{eq:dS}) shows the mid price of the underlying asset is Brownian motion with volatility.
  \item Equation (\ref{eq:dQ}) represents that the remaining inventory decreases 1 unit when LO or MM matches.
\end{itemize}
The equations (\ref{eq:X})-(\ref{eq:Q}) represents the market order as th impulse.
\begin{itemize}
  \item Equation (\ref{eq:X}) represents the increase in cash when the agent executes market order of amount $\zeta_i$ at $\tau_i$.
  Since the market order is executed at the bid price , there is an increase in cash of $(S_{\tau_i}-\xi)\zeta_i$.
  Additionally, market impact is considered in the form of a power of the amount $\alpha_M \zeta_i^{{\beta_M}}$.
\end{itemize}
Let $\mathcal{A}(s, y)$ be a class of combined stochastic control $\delta=(\delta^L, \delta^I): [s, T]\times \Omega \rightarrow {\bf R}^2$ and impulse controls $\theta$ such that 
existence and uniqueness of the SDEs (\ref{eq:dX})-(\ref{eq:dQ}) be guaranteed. 
Then the value function is 
\begin{align}
  H(t,x,s, q) = \sup_{(\delta, \theta) \in \mathcal{A}(t, y)} E\left[ X_T + Q_T(S_T-\xi -\alpha Q_T) -\alpha_M Q_T^{{\beta_M}} - \phi \int_t^T (Q_s-\bar{q}_s)^2 ds \right]. \label{eq:val}
\end{align}
The value function (\ref{eq:val}) is composed expectation of the following three terms.
\begin{enumerate}
  \item The first term represents the terminal cash $X_T$
  \item The second term represents that if terminal inventory $Q_T > 0$, then the agenet should sell all remaing amount through market order at $Q_T(S_T-\xi)$, and we assume it cause market impact $\alpha Q_T$
  \item The third term $\phi \int_t^T (Q_s-\bar{q}_s)^2ds$ shows the penalty for deviation from the benchmark schedule. $\bar{q}$ shows the benchmark e.g. static strategy of Almgren-Chriss \cite{almgren2001optimal} can be used.
\end{enumerate}

\subsection{The Resulting HJB QVI}
The infinitesimal generator $L^{\delta}$ of the state processes driven by (\ref{eq:dX})-(\ref{eq:dQ}) is
\begin{align*}
   &L^{\delta} H = \frac{1}{2}\sigma^2\frac{\partial^2H}{\partial s^2} \\
  &+ [H(t, x+ s+\delta^L-\alpha^L\lambda_L e^{-\kappa \delta^L} , s, q -1) - H(t, x, s, q)]\lambda_L e^{-\kappa \delta^L} \\
  &+[H(t, x+ s+\delta^I, s, q-1 ) - H(t, x, s, q)]\lambda_I e^{-\kappa \delta^I} .
\end{align*}
The impulse operator driven by (\ref{eq:X})-(\ref{eq:Q}) is
$$\mathcal{M}H(t,x,s, q) = \max_{\zeta\in \{0, 1, \cdots, q\}} H(t,x + (s-\xi)\zeta-\alpha_M\zeta^{{\beta_M}},s, q-\zeta). $$
The integrand $f$ and terminal function $g$ in the equation (\ref{eq:valGen}) are corresponding to
$$f(t, x,s,q,\delta) = -\phi  (q-\bar{q}_t)^2, g(x, s, q) = x + q(s-\xi-\alpha q) -\alpha_M q^{{\beta_M}}.$$
By applying these setup to the general framework in the previous section \ref{sec:math}, 
the value function (\ref{eq:val}) is characterized as the unique viscosity solution of the following HJBQVI.
\begin{align} \label{eq:HJBQVI2}
    &\max\left\{\frac{\partial H}{\partial t} + \frac{1}{2}\sigma^2\frac{\partial^2H}{\partial s^2} - \phi  (q-\bar{q}_t)^2\right.\nonumber\\
    &\left. + \sup_{\delta^L, \delta^I}([H(t, x+ s+\delta^L-\alpha^L\lambda_L e^{-\kappa \delta^L} , s, q -1) - H(t, x, s, q)]\lambda_L e^{-\kappa \delta^L} \right. \nonumber\\
    &\left. + [H(t, x+ s+\delta^I, s, q-1 ) - H(t, x, s, q)]\lambda_I e^{-\kappa \delta^I} ), \right. \nonumber\\
    &\left.  \max_{\zeta\in \{0, 1, \cdots, q\}} H(t,x+(s-\xi)\zeta-\alpha_M\zeta^{{\beta_M}},s, q-\zeta) - H(t,x,s,q)\right\}  = 0, 
\end{align}
with boundary and terminal conditions
\begin{align*}
  H(t,x,s,0) &= x -\phi \int_t^T \bar{q_s}^2 ds,\\
  H(T,x,s,q) &= x + q(s - \xi - \alpha q) -\alpha_M q^{{\beta_M}}.
\end{align*}.
\subsection{Solving the equation}
From terminal and boundary conditions, we assume the ansatz for the value function
$H(t, x, s, q) = x + qs + h(t, q).$ Notice that $\sup_{\delta^L, \delta^I}$ in the HJB QVI(\ref{eq:HJBQVI2}) can be separated as
\begin{align*}
    \sup_{\delta^L, \delta^I}\left([H(t, x+ s+\delta^L-\alpha^L\lambda_L e^{-\kappa \delta^L} , s, q -1) - H(t, x, s, q)]\lambda_L e^{-\kappa \delta^L}  \right.\\
    \left.+ [H(t, x+ s+\delta^I, s, q-1 ) - H(t, x, s, q)]\lambda_I e^{-\kappa \delta^I}   \right)\\
    =\sup_{\delta^L}\left([H(t, x+ s+\delta^L-\alpha^L\lambda_L e^{-\kappa \delta^L} , s, q -1) - H(t, x, s, q)]\lambda_L e^{-\kappa \delta^L} \right)  \\
    + \sup_{\delta^I}\left([H(t, x+ s+\delta^I, s, q-1 ) - H(t, x, s, q)]\lambda_I e^{-\kappa \delta^I}  \right), 
\end{align*}
Then HJB QVI (\ref{eq:HJBQVI2}) reduces considarably to the following equation for $h(t, q)$
\begin{align}
  &\max\left\{\partial_t h(t,q)-\phi (q-\bar{q}_t)^2 h(t,q) +e^{-1}\lambda_I \omega(t,q-1)\right.\label{eq:h1}\\
  &\left.+\sup_{\delta_L}  \left(\delta_L - \alpha_L\lambda_L e^{-\kappa \delta_L}+h(t, q-1)-h(t, q) \right)\lambda_L e^{-\kappa \delta_L} \right. \label{eq:h2}\\
  &\left.+\sup_{\delta_I} \left(\delta_I -  h(t, q-1)-h(t, q) \right)\lambda_I e^{-\kappa \delta_I} , \right.\label{eq:h3}\\
  &\left. \max_{\zeta\in \{0, 1, \cdots, q\}} \left(h(t, q-\zeta) - h(t, q) - (\xi \zeta + \alpha_M\zeta^{{\beta_M}} )\right) \right\} =0, \label{eq:h4}
\end{align}
with boundary and terminal conditions
\begin{align*}
  h(t,0) &=  -\phi \int_t^T \bar{q_s}^2 ds,\\
  h(T,q) &= - q(\xi + \alpha q)-\alpha_M q^{{\beta_M}}.
\end{align*}
Focusing on the term (\ref{eq:h3}), the optimal MM spread $\delta^{I*}_t$ is explicitly obtained as
\begin{align}
  \delta^{I*}_t = \frac{1}{\kappa} + h(t, q) - h(t, q-1). \label{eq:delta^I}
\end{align}
Focusing on the term (\ref{eq:h3}), the optimal LO depth $\delta^{L*}_t$ is obtained as the solution of the following equation.
An analytical expression is hard to derive, however, it can be computed numerically without difficulty using methods such as the Newton-Raphson method or the bisection method.
\begin{align}
  1-\kappa \delta^{L*}_t + 2 \kappa \alpha_L \lambda_L e^{-\kappa \delta^{L*}_t} -\kappa(h(t, q-1)-h(t, q)) = 0.
\end{align}
Focusing on the term (\ref{eq:h4}), it is similarly difficult to analytically derive the optimal MO size $\zeta^*$. 
However, it is possible to numerically determine the $\zeta^*$ that attains the maximum by ultimately employing a brute-force search method.
The timing of MO execution is 
\begin{align}
  \tau_i = \inf\left\{t > \tau_{i-1};  h(t, q-\zeta^*) - h(t, q) = \xi \zeta^* + \alpha_M(\zeta^*)^{{\beta_M}} \right\},\label{eq: tau}
\end{align}
for $i = 1, 2, \cdots.$

Based on these considerations, the HJB QVI can be numerically solved using PDE numerical computation methods, e.g., the Finite Difference Method (FDM).
Once $h(t, q)$ is obtained, we can also derive following important values for optimal strategy.
\begin{itemize}
  \item $\delta^{I*}_t$: The optimal ask spread at which the agent shows clients through the MM is obtained from the equation (\ref{eq:delta^I}).
  \item $\delta^{L*}_t$: The optimal depth at which the agent posts LO is obtained from (\ref{eq:h3})
  \item $\tau_i$: The optimal timing timing of MO execution is obtained from the equation (\ref{eq: tau}).
\end{itemize}

\section{Numerical Experiments}
We carry out a numerical implementation to compute the optimal execution strategy,
including the execution timiming of MO, the depth of LOs, and ask MM spread.  
We compare the following strategies.
\begin{enumerate}
  \item ACAlmgren-Chriss model \cite{almgren2001optimal}. It is a static strategy $\bar{q}_t$, 
  \begin{align}
    \bar{q}_t = Q_0\frac{\sinh(\gamma (T-t))}{\sinh(\gamma T)}.\label{eq:alm}
  \end{align}
  \item LO/MO model by Cartea et al. \cite{cartea2015algorithmic}, which includes benchmark (\ref{eq:alm}) as a benchmark.
  \item LO/MO/MM model proposed in section\ref{sec:model}, which includes benchmark (\ref{eq:alm}) as a benchmark in the equation (\ref{eq:val}).
\end{enumerate}
Our implementation is found at the following link.\\
\url{https://gist.github.com/morimoto0606/20a199e41408455913ef91406b6d6861}\\
It is based on the sample code by Sebastian Jaimungal.\\
\url{https://sebastian.statistics.utoronto.ca/books/algo-and-hf-trading/code/}\\
The parameters are as follows.
\begin{align*}
T = 60, Q_0 = 10, \sigma = 0.01, \kappa_0 = 0.1, \xi = 0.01,  \alpha = 0.0001, \phi = 0.001, \gamma = 0.1\\
\kappa = 100, \lambda_L = 50/60, \lambda_I = 60/60, \alpha_L = 0.005, \alpha_M = 0.0001, {\beta_M} = 1.5
\end{align*}

\begin{figure}[H]
  \centering
  \includegraphics[width=0.9\textwidth]{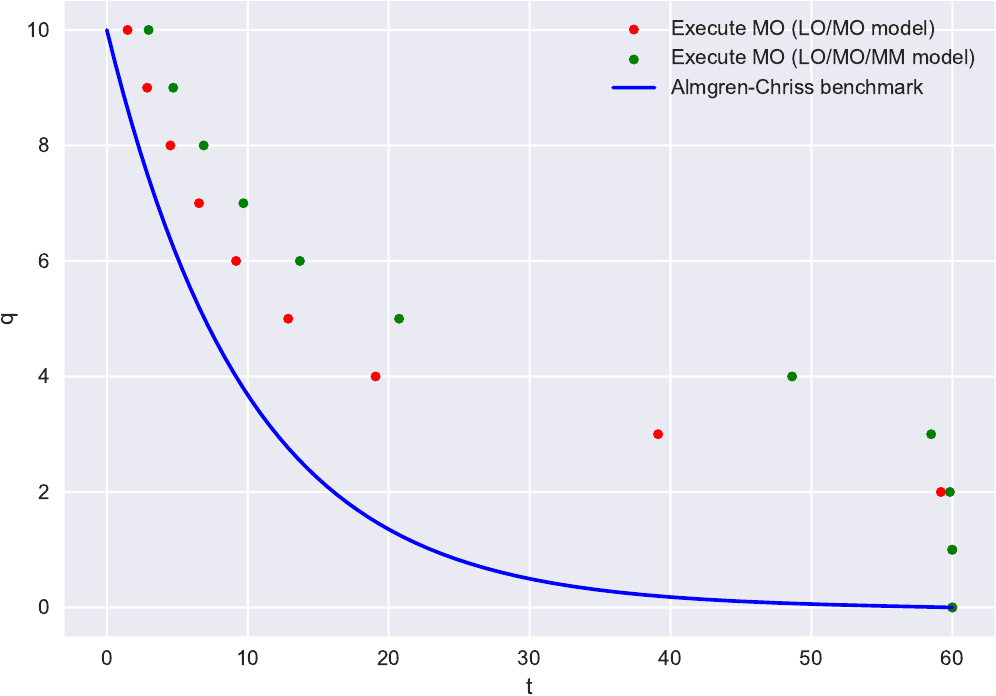}
  \caption{Optimal execution schdule and MO timing}
  \label{fig:mo}  
\end{figure}

Figure \ref{fig:mo} shows Figure 1 illustrates the execution schedule by MO in the case where neither LO nor MM are filled for each strategy. 
The blue line represents Strategy 1 Almgren-Chriss benchmark, which corresponds to a static execution schedule used as a benchmark for Strategy 2 and 3.
The red and green dots represent the times and inventories when MO are executed for Strategy 2 and 3, respectively. 
As for the execution methods, Strategy 2 uses both MO and LO, while Strategy 3 also incorporates internal orders, utilizing MM to further enhance profitability. 
As a result, Strategy 2 is behind the benchmark schedule of Strategy 1 due to the use of LO in addition to MO, and Strategy 3 is further delayed compared to Strategy 2, 
as it waits for internal orders through MM.
Table \ref{tab:MO} shows the actual time of red dot $\tau_L$ and green dot $\tau_I$.

\begin{table}[H]
  \centering
  \caption{Optimal MO execution timing}
  \label{tab:MO}
  \begin{tabular}{c|ccccccccccc} 
  $q$ & 10 & 9 & 8 & 7 & 6 & 5 & 4 & 3 & 2 & 1 & 0 \\
  \hline
  $\tau_{\text{LO/MO}}$ & 1.47 & 2.86 & 4.51 & 6.54 & 9.17 & 12.87 & 19.07 & 39.12 & 59.19 & 59.99 & 60.00 \\
  $\tau_{\text{LO/MO/MM}}$ & 2.96 & 4.71 &	6.87 & 9.69	& 13.70	& 20.74	& 48.63	& 58.50	& 59.83	& 60.00	& 60.00 \\
  \end{tabular}
\end{table}

Table \ref{tab:zeta} shows the optimal market order execution size. 
Notice that in the LO/MO/MM model, the market order size 
$\zeta$ was also chosen optimally. 
$\zeta$ is a function of $t$ and $q$, and here it represents the statistics for the sample of 
$q=0$ to $10$ at each $t=10 - 50$. 
According to this, there is a trend that the execution size of the market order increases as time progresses. 
This can be attributed to the increasing need to complete the execution using the MO, 
even if the price is not as favorable as the LO or MM, due to the diminishing time remaining. 
Since market impact is incorporated as the $0.5$ power of the execution size, 
it is considered more efficient to execute 2 units in one MO rather than executing 1 unit twice."
\begin{table}[H]
  \centering
  \caption{Statistics of Optimal market order size of LO/MO/MM model}\label{tab:zeta}
  \begin{tabular}{|c|c|c|c|c|c|c|c|c|}
  \hline
  Time & count & mean & std & min & 25\% & 50\% & 75\% & max \\
  \hline
  10	& 11.0	& 0.545455	& 0.687552	& 0.0	& 0.0	& 0.0	& 1.0	& 2.0 \\
  20	& 11.0	& 0.909091	& 1.044466	& 0.0	& 0.0	& 1.0	& 1.5	& 3.0 \\
  30	& 11.0	& 1.272727	& 1.348400	& 0.0	& 0.0	& 1.0	& 2.0	& 4.0 \\
  40	& 11.0	& 1.272727	& 1.348400	& 0.0	& 0.0	& 1.0	& 2.0	& 4.0 \\
  50	& 11.0	& 1.545455	& 1.368476	& 0.0	& 0.5	& 1.0	& 2.5	& 4.0 \\
  \hline
  \end{tabular}
\end{table}

\begin{figure}[H]
  \centering
  \includegraphics[width=0.9\textwidth]{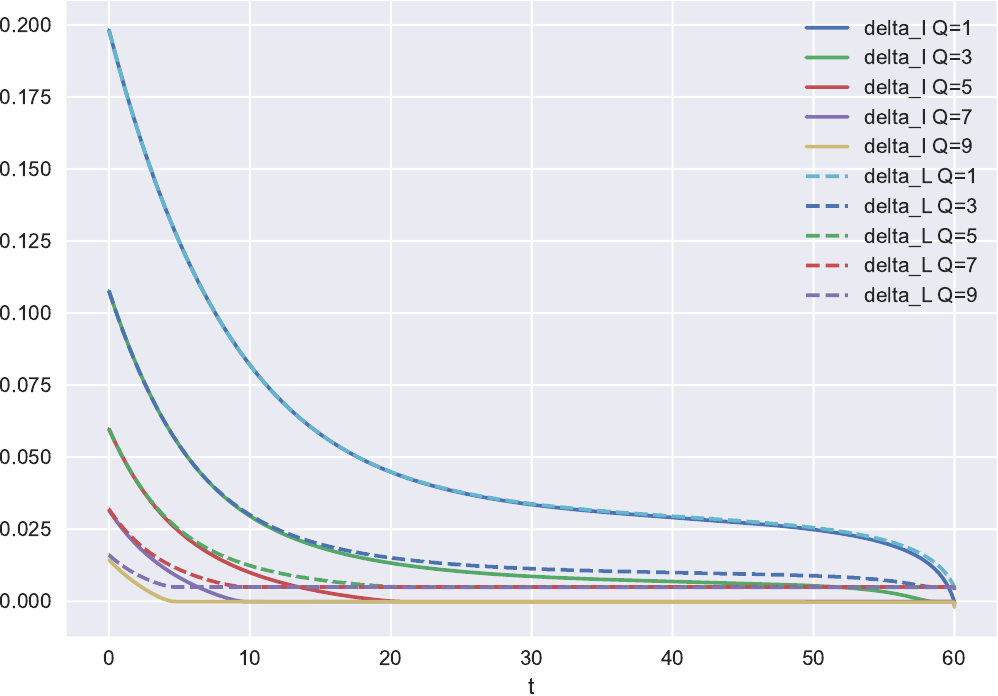}
  \caption{Optimal spread $\delta^I$ and depth $\delta^L$}
  \label{fig:delta}  
\end{figure}

Table\ref{tab:delta_l}, \ref{tab:delta_i} and Figure \ref{fig:delta} shows, for Strategy 3, the optimal LO depth (dashed line) for each inventory at each time, 
as well as the optimal ask MM apread (solid line).
For simplicity in presentation, only the curves for odd inventory levels are shown here.
When comparing the LO depth $\delta^L$ and MM spread $\delta^I$ for the same inventory, 
they are nearly identical when there is still plenty of time remaining and the inventory is small around 1. 
On the other hand, when the remaining time is short and there is more inventory left, the MM spread is thinner than the LO depth (client-favorable) price, and it is particularly suggested that a better price can be offered, even beyond the mid.
This suggests that there is no market impact during execution through MM, and since execution can occur at a better price than the MO, even beyond the mid, this behavior aligns well with real-world experience.

\begin{table}[H]
  \centering
  \caption{LO depth}
  \label{tab:delta_l}
  \begin{tabular}{lccccc} 
  \hline
  $t$ & $\delta_t^{L*}(q=1)$ & $\delta_t^L(q=3)$ & $\delta_t^L(q=5)$ & $\delta_t^L(q=7)$ & $\delta_t^L(q=9)$ \\
  \hline
  10& 0.081970	& 0.029968	& 0.012390	& 0.004944	& 0.004944 \\
  20& 0.044861	& 0.015076	& 0.005066	& 0.004944	& 0.004944 \\
  30& 0.033752	& 0.011292	& 0.004944	& 0.004944	& 0.004822 \\
  40& 0.029480	& 0.009949	& 0.004944	& 0.004944	& 0.004822 \\
  50& 0.025574	& 0.008850	& 0.004944	& 0.004944	& 0.004822 \\
  \hline
  \end{tabular}
\end{table}

\begin{table}[H]
  \centering
  \caption{MM spread}
  \label{tab:delta_i}
  \begin{tabular}{lccccc} 
  \hline
  $t$ & $\delta_t^I(q=1)$ & $\delta_t^I(q=3)$ & $\delta_t^I(q=5)$ & $\delta_t^I(q=7)$ & $\delta_t^I(q=9)$ \\
  \hline
  10& 0.081992	& 0.029569	& 0.009957	&-0.000100  &-0.000100 \\
  20& 0.044826	& 0.013250	& 0.000050	&-0.000100  &-0.000100 \\
  30& 0.033527	& 0.008625	& -0.00010  &-0.000183 &-0.000237 \\
  40& 0.029097	& 0.006878	& -0.00010  &-0.000183 &-0.000237 \\
  50& 0.024879	& 0.005403	& -0.00010  &-0.000100 &-0.000280 \\
  \hline
  \end{tabular}
\end{table}

\section*{Disclosure of interest}
The author declares that there are no competing interests.

\section*{Funding}
No funding was received for this research.

\section*{Acknowledgments}
The author would like to thank Dr. Keiichi Maeta of MUFG Bank for insightful discussions, and Mr. Katsumi Takada of Diva Analytics for his valuable comments.

\bibliographystyle{plain}
\bibliography{references}  

\end{document}